\documentclass[aps,preprint,showpacs]{revtex4}
\usepackage{amsfonts}
\usepackage{amsmath}
\usepackage{amssymb}
\usepackage{graphicx}
\begin{document}

\title{Effect of the variation of mass on fermion localization on thick branes}
\author{Luis B. Castro}\email[ ]{benito@feg.unesp.br}\author{Luis A. Meza}\email[ ]{luisarroyo@feg.unesp.br}
\affiliation{Departamento de F\'{\i}sica e Qu\'{\i}mica,
Universidade Estadual Paulista, 12516-410 Guaratinguet\'{a}, S\~ao
Paulo, Brazil} 


\begin{abstract}

In a recent paper published in this journal, Zhao and collaborators [Phys. Rev. D \textbf{82}, 084030 (2010)] analyze a toy model of thick branes generated by a real scalar field with the potential $V(\phi)=a\phi^{2}-b\phi^{4}+c\phi^{6}$, and investigate the variation of the mass parameter $a$ on the branes as well as the localization and resonances of fermions. In that research the simplest Yukawa coupling $\eta\bar{\Psi}\phi\Psi$ was considered. In that work does not analyze the zero mode in details and also contains some misconceptions. In this paper, the effect of the variation of the mass parameter $a$ on the brane is reinvestigated and it is associated to the phenomenon of brane splitting. Furthermore, it is shown that the zero mode for the left-handed fermions can be localized on the brane depending on the values for the coupling constant $\eta$ and the mass parameter $a$.
\end{abstract}

\maketitle

\section{Introduction}

The authors of Ref. \cite{zhao} investigated the effects of the variation of the mass parameter $a$ on the thick branes. They used a real scalar field, which has a potential of the $\phi^{6}$ model, as the background field of the thick branes. It was found that the number of the bound states (in the case without gravity) or the resonant states (in the case with gravity) increases with the parameter $a$. That work considered the simplest Yukawa coupling $\eta\bar{\Psi}\phi\Psi$, where $\eta$ is the coupling constant. The authors stated that as the value of $a$ is increasing, the maximum of the matter energy density splits into two new maxima, and the distance of the new maxima increases and the brane gets thicker. The authors also stated that the brane with a big value of $a$ would trap fermions more efficiently.

In this paper, we reinvestigated the effect of the variation of the mass parameter $a$ on the thick branes, because the above investigation does not analyze the zero mode in details and contains some misconceptions. We only focus attention in the case with gravity. We find that the variation of $a$ on the thick brane is associated to the phenomenon of brane splitting. From the static equation of motion, we analyze the asymptotic behavior of $A(y)$ and find that the zero mode for left-handed fermions can be localized on the brane depending on the value for the coupling constant $\eta$ and the mass parameter $a$. We also show that as the value of $a$ is increasing the simplest Yukawa coupling does not support the localization of fermions on the brane, as incompletely argued in Ref. \cite{zhao}.

\section{Thick brane with gravity}

The action for our system is described by \cite{cam}
\begin{equation}\label{ac1}
    S=\int d^4xdy\sqrt{ -g}\left[\frac{1}{4}\,R-\Lambda-\frac{1}{2}g^{MN}\partial_{M}\phi\partial^{N}\phi-V(\phi) \right],
\end{equation}
\noindent where $M,N=0,1,2,3,4$, $\Lambda$ is the 5D bulk cosmological constant and the scalar potential $V(\phi)$ is given by \cite{zhao}
\begin{equation}\label{pot1}
    V(\phi)=a\phi^{2}-b\phi^{4}+c\phi^{6}\,,
\end{equation}

\noindent where $a,b,c>0$. There are three minima for $V(\phi)$, one is at $\phi^{(1)}=0$ (local minima) corresponding to a disordered bulk phase and the other two are at $\phi^{(2)}=-\phi^{(3)}=v$ (global minima) with
\begin{equation}\label{v}
    v=\sqrt{\frac{\sqrt{b^{2}-3ac}}{3c}+\frac{b}{3c}}\,.
\end{equation}

\noindent They are degenerated and correspond to ordered bulk phases. As $a=a_{c}$ ($a_{c}=b^{2}/4c$), $V(\phi^{(1)})=V(\phi^{(2)})=V(\phi^{(3)})$, $V(\phi)$ has three degenerated global minima. For the case with gravity, the critical value of $a$ is not $a_{c}$ but a smaller effective critical value $a_{*}$. In this case, $a_{c}=a_{*}=0.837$ \cite{zhao}. The line element in this model is considered as
\begin{equation}\label{metric}
    ds^{2}=g_{ab}dx^{a}dx^{b}=\mathrm{e}^{2A(y)}\eta_{\mu\nu}dx^{\mu}dx^{\nu}+dy^{2},
\end{equation}

\noindent where $\mu,\nu=0,1,2,3$, $\eta_{\mu\nu}=\mathrm{diag}(-1,1,1,1)$, and $\mathrm{e}^{2A}$ is the so-called warp factor. We suppose that $A=A(y)$ and $\phi=\phi(y)$.

\subsection{Effects of the variation of the mass parameter $a$ on the thick brane}

For this model, the equations of motion are
\begin{equation}\label{em1b}
    \phi^{\prime\prime}=-4A^{\prime}\phi^{\prime}+\frac{dV(\phi)}{d\phi},
\end{equation}
\begin{equation}\label{em2b}
    A^{\prime\prime}+2A^{\prime}\,^{2}=-\frac{1}{3}\,\phi^{\prime}\,^{2}-%
    \frac{2}{3}\,(V+\Lambda)\,,
\end{equation}
\begin{equation}\label{em3b}
    A^{\prime}\,^{2}=\frac{1}{6}\,\phi^{\prime}\,^{2}-\frac{1}{3}\,(V+\Lambda)\,.
\end{equation}

\noindent It is possible to rewrite (\ref{em2b}) and (\ref{em3b}) as
\begin{equation}\label{em3c}
A^{\prime\prime}=-\frac{2}{3}\,\phi^{\prime}\,^{2}\,.
\end{equation}

\noindent The boundary conditions can be read as follows
\begin{equation}\label{bc1}
    A(0)=A^{\prime}(0)=\phi(0)=0\,,
\end{equation}
\begin{equation}\label{bc2}
    \phi(+\infty)=-\phi(-\infty)=v.
\end{equation}

The matter energy density has the form
\begin{equation}\label{de}
    T_{00}=\rho(y)=\mathrm{e}^{2A(y)}\left[ \frac{1}{2}\,\left( \frac{d\phi}{dy}%
    \right)^{2}+V\left(\phi\right) \right].
\end{equation}

\noindent At this point, it is also instructive to analyze the matter energy of the toy model
\begin{equation}\label{ephi}
    E_{\phi}=\int^{\infty}_{-\infty}dy\,T_{00}\,,
\end{equation}

\noindent substituting (\ref{de}) in (\ref{ephi}), we get
\begin{equation}\label{ephi2}
    E_{\phi}=\int^{\infty}_{-\infty}dy\,\mathrm{e}^{2A(y)}\left[ \frac{1}{2}\,\left( \frac{d\phi}{dy}%
    \right)^{2}+V\left(\phi\right) \right]\,,
\end{equation}

\noindent using (\ref{em3b}) and (\ref{em3c}), we obtain the value of the matter energy given by
\begin{equation}\label{ephi3}
    E_{\phi}=\frac{3}{2}\,\left[\mathrm{e}^{2A(-\infty)}A^{\prime}(-\infty)-%
    \mathrm{e}^{2A(\infty)}A^{\prime}(\infty) \right]-\Lambda\int^{\infty}_{-\infty}dy%
    \mathrm{e}^{2A(y)}.
\end{equation}

\noindent As $\Lambda=0$, the value of the matter energy depends on the asymptotic behavior of the warp factor. If $y\rightarrow\pm\infty$ then $\phi^{\prime}(\pm\infty)=0$ and by the analysis to Eq. (\ref{em2b}), we can see that $A(\pm\infty)\propto -|\,y|$. Therefore, $\mathrm{e}^{2A(\pm\infty)}\rightarrow0$ and the value of the matter energy is zero. This fact is the same to the case of branes with generalized dynamics \cite{arro}.

\noindent The scalar curvature (or Ricci scalar) is given by
\begin{equation}\label{ricci}
    R=-4(5A^{\prime}\,^{2}+2A^{\prime\prime}).
\end{equation}

\begin{figure}[ht]
\begin{center}
\includegraphics[width=7cm, angle=0]{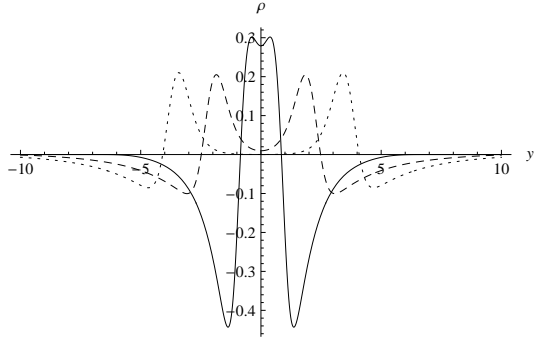}
\end{center}
\caption{The profiles of the energy density for $b=2$, $c=1$, $a=0$ (thin line), $a=0.8$ (dashed line) and $a=0.836$ (dotted line).} \label{fde}
\end{figure}

\begin{figure}[ht]
\begin{center}
\includegraphics[width=7cm, angle=0]{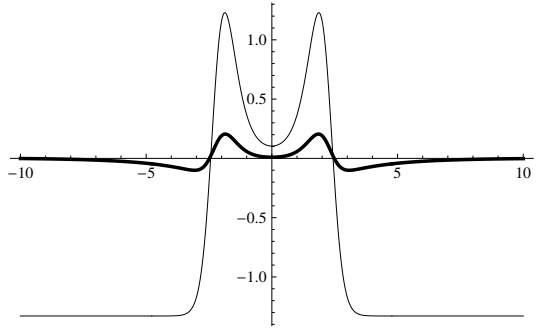}
\end{center}
\caption{The profiles of the matter energy density (thin line) and Ricci scalar (thick line) for $b=2$, $c=1$ and $a=0.8$.} \label{desc}
\end{figure}

\noindent The profiles of the matter energy density is shown in Fig. (\ref{fde}) for some values of $a$. Figure (\ref{fde}) clearly shows that for $a=0$ the matter energy density has not a single-peak around $y=0$. The core of the brane is localized at $y=0$ for $a=0$, because this region has a positive matter energy density. On the other hand, as the value of $a$ is increasing, we can see that the single brane splits into two sub-branes and as $a\rightarrow a_{*}$ each sub-brane is a thick brane. This phenomenon is so-called of brane splitting \cite{angel}. From the peak of the matter energy density is evident know where the core of the branes are located. Therefore, the brane does not get thicker with the increases of the value of the mass parameter $a$, as argued in Ref. \cite{zhao}. The profiles of the matter energy density and the Ricci scalar are shown in Fig. (\ref{desc}) for $a=0.8$. Note that the presence of regions with positive Ricci scalar is connected to the capability to trap matter near to the core of the brane \cite{alm} and it reinforces the conclusion of the analyzes from the matter energy density. Also note that far from the brane, $R$ tends to a negative constant, characterizing the $AdS_{5}$ limit from the bulk.

\subsection{Fermion localization}

The action for a Dirac spinor field coupled with the scalar fields by a general Yukawa coupling is
\begin{equation}\label{ad}
    S=\int d^{5}x\sqrt{|\,g|}\left[ i\bar{\Psi}\Gamma^{M}\nabla_{M}\Psi-\eta\bar{\Psi}F(\phi)\Psi \right]\,,
\end{equation}
\noindent where $\eta$ is the positive coupling constant between fermions and the scalar field. Moreover,
we are considering the covariant derivative $\nabla_{M}=\partial_{M}+\frac{1}{4}\,\omega^{\bar{A}\bar{B}}_{M}\Gamma_{\bar{A}}\Gamma_{\bar{B}}$\,,
where $\bar{A}$ and $\bar{B}$, denote the local Lorentz indices and $\omega^{\bar{A}\bar{B}}_{M}$ is
the spin connection. Here we consider the field $\phi$ as a background field. The equation of motion is
obtained as
\begin{equation}\label{dkp}
i\,\Gamma ^{M }\nabla_{M}\Psi-\eta F(\phi)\Psi =0.
\end{equation}%

At this stage, it is useful to consider the fermionic current. The conservation law for $J^{M}$ follows from the standard procedure and it becomes
\begin{equation}\label{corr}
\nabla_{M}J^{M}=\bar{\Psi}\left(\nabla_{M}\Gamma^{M}\right)\Psi\,,
\end{equation}
\noindent where $J^{M}=\bar{\Psi}\Gamma^{M}\Psi$. Thus, if
\begin{equation}\label{cj0}
    \nabla_{M}\Gamma^{M}=0\,,
\end{equation}
\noindent then four-current will be conserved. The condition (\ref{cj0}) is the purely geometrical assertion that the curved-space gamma matrices are covariantly constant.

\noindent Using the same line element (\ref{metric}) and the representation for
gamma matrices $\Gamma^{M}=\left( \mathrm{e}^{-A}\gamma^{\mu},-i\gamma^{5}\right)$,
the condition (\ref{cj0}) is trivially satisfied and therefore the current is conserved.

The equation of motion (\ref{dkp}) becomes
\begin{equation}\label{em}
\left[ i\gamma^{\mu}\partial_{\mu}+\gamma^{5}\mathrm{e}^{A}(\partial_{y}+2\partial_{y}A)%
-\eta\,\mathrm{e}^{A}F(\phi) \right]\Psi=0.
\end{equation}
\noindent Now, we use the general chiral decomposition
\begin{equation}\label{dchiral}
    \Psi(x,y)=\sum_{n}\psi_{L_{n}}(x)\alpha_{L_{n}}(y)+\sum_{n}\psi_{R_{n}}(x)\alpha_{R_{n}}(y),
\end{equation}
\noindent with $\psi_{L_{n}}(x)=-\gamma^{5}\psi_{L_{n}}(x)$ and $\psi_{R_{n}}(x)=\gamma^{5}\psi_{R_{n}}(x)$.
With this decomposition $\psi_{L_{n}}(x)$ and $\psi_{R_{n}}(x)$ are the left-handed and
right-handed components of the four-dimensional spinor field, respectively. After applying
(\ref{dchiral}) in (\ref{em}), and demanding that $i\gamma^{\mu}\partial_{\mu}\psi_{L_{n}}=m_{n}\psi_{R_{n}}$
and $i\gamma^{\mu}\partial_{\mu}\psi_{R_{n}}=m_{n}\psi_{L_{n}}$, we obtain two equations
for $\alpha_{L_{n}}$ and $\alpha_{R_{n}}$
\begin{equation}\label{ea1}
    \left[ \partial_{y}+2\partial_{y}A+\eta F(\phi) \right]\alpha_{L_{n}}=m_{n}\mathrm{e}^{-A}\alpha_{R_{n}}\,,
\end{equation}
\begin{equation}\label{ea2}
    \left[ \partial_{y}+2\partial_{y}A-\eta F(\phi) \right]\alpha_{R_{n}}=-m_{n}\mathrm{e}^{-A}\alpha_{L_{n}}\,.
\end{equation}
\noindent Inserting the general chiral decomposition (\ref{dchiral}) into the action (\ref{ad}), using (\ref{ea1}) and (\ref{ea2}) and also requiring that the result take the form of the standard four-dimensional action for the massive chiral fermions
\begin{equation}\label{ad2}
    S=\sum_{n}\int d^{4}x\, \bar{\psi}_{n}\left( \gamma^{\mu}\partial_{\mu}-m_{n} \right)\psi_{n},
\end{equation}
\noindent where $\psi_{n}=\psi_{L_{n}}+\psi_{R_{n}}$ and $m_{n}\ge0$, the functions $\alpha_{L_{n}}$ and $\alpha_{R_{n}}$ must obey the following orthonormality conditions
\begin{equation}\label{orto}
    \int_{-\infty}^{\infty}dy\,\mathrm{e}^{3A}\alpha_{Lm}\alpha_{Rn}=\delta_{LR}\delta_{mn}.
\end{equation}

\noindent Implementing the change of variables %
\begin{equation}\label{cv}
    z=\int^{y}_{0}\mathrm{e}^{-A(y^{\,\prime})}dy^{\,\prime},
\end{equation}

\noindent $\alpha_{L_{n}}=\mathrm{e}^{-2A}L_{n}$ and $\alpha_{R_{n}}=\mathrm{e}^{-2A}R_{n}$, we get
\begin{equation}\label{sleft}
    -L_{n}^{\prime\prime}(z)+V_{L}(z)L_{n}=m_{n}^{2}L_{n}\,,
\end{equation}
\begin{equation}\label{sright}
    -R_{n}^{\prime\prime}(z)+V_{L}(z)R_{n}=m_{n}^{2}R_{n}\,,
\end{equation}
\noindent where
\begin{eqnarray}
  V_{L}(z) &=& \eta^{2}\mathrm{e}^{2A}F^{2}(\phi)-\eta\partial_{z}\left( \mathrm{e}^{A}F(\phi) \right),\label{vefa} \\
  V_{R}(z) &=& \eta^{2}\mathrm{e}^{2A}F^{2}(\phi)+\eta\partial_{z}\left( \mathrm{e}^{A}F(\phi) \right)\label{vefb}.
\end{eqnarray}
\noindent Using the expressions $\partial_{z}A=\mathrm{e}^{A(y)}\partial_{y}A$ and $\partial_{z}F=\mathrm{e}^{A(y)}\partial_{y}F$, we can recast the potentials (\ref{vefa}) and (\ref{vefb}) as a function of $y$ \cite{yu}-\cite{yo}
\begin{eqnarray}
  V_{L}(z(y)) &=& \eta\mathrm{e}^{2A}\left[ \eta F^{2}-\partial_{y}F-F\partial_{y}A(y) \right],\label{vya} \\
  V_{R}(z(y)) &=& V_{L}(z(y))|_{\eta\rightarrow-\eta}\,.\label{vyb}
\end{eqnarray}

\noindent It is worthwhile to note that we can construct the Schr\"{o}dinger potentials
$V_{L}$ and $V_{R}$ from eqs. (\ref{vya}) and (\ref{vyb}).

At this stage, it is instructive to state that with the change of variable (\ref{cv})
we get a geometry to be conformally flat
\begin{equation}\label{cm}
    ds^{2}=\mathrm{e}^{2A(z)}\left( \eta_{\mu\nu}dx^{\mu}dx^{\nu} +dz^{2}\right).
\end{equation}
\noindent Now we focus attention on the condition (\ref{cj0}) for the line element (\ref{cm}). In this case we obtain
\begin{equation}\label{mc}
    \nabla_{M}\Gamma^{M}= i(\partial_{z}A(z))\mathrm{e}^{-A(z)}\gamma^{5}.
\end{equation}

\noindent Therefore, the current is no longer conserved for the line element (\ref{cm}) \cite{arro}. It is known that, in general, the reformulation of the theory in a new conformal frame leads to a different, physically inequivalent theory. This issue has already a precedent in cosmological models \cite{val}.

\noindent Under this arguments, we only use the change of variable (\ref{cv}) to have a qualitative analysis of the potential profiles (\ref{vya}) and (\ref{vyb}), which is a fundamental ingredient for the fermion localization on the brane.

Now we focus attention on the calculation of the zero mode. Substituting $m_{n}=0$ in (\ref{ea1}) and (\ref{ea2}) and using $\alpha_{L_{n}}=\mathrm{e}^{-2A}L_{n}$ and $\alpha_{R_{n}}=\mathrm{e}^{-2A}R_{n}$, respectively, we get
\begin{equation}\label{mzL}
    L_{0}\propto \exp \left[-\eta\int_{0}^{y}dy^{\prime}F(\phi) \right],
\end{equation}
\begin{equation}\label{mzR}
    R_{0}\propto \exp \left[\eta\int_{0}^{y}dy^{\prime}F(\phi) \right].
\end{equation}
\noindent This fact is the same to the case of two-dimensional Dirac equation \cite{luis}. At this point is worthwhile to mention that the normalization of the zero mode and the existence of a minimum for the effective potential at the localization on the brane are essential conditions for the problem of fermion localization on the brane. This fact was already reported in \cite{yo}.

In order to guarantee the normalization condition (\ref{orto}) for the left-handed fermion zero mode (\ref{mzL}), the integral must be convergent, \textit{i.e}
\begin{equation}\label{cono}
    \int^{\infty}_{-\infty}dy\exp\left[ -A(y)-2\eta\int^{y}_{0}dy\,^{\prime}F(\phi(y\,^{\prime})) \right]<\infty.
\end{equation}
\noindent This result clearly shows that the normalization of the zero mode is decided by the asymptotic behavior of $F(\phi(y))$. Furthermore, from (\ref{vya}) and (\ref{vyb}), it can be observed that the effective potential profile depends on the $F(\phi(y))$ choice. This fact implies that the existence of a minimum for the effective potential $V_{L}(z(y))$ or $V_{R}(z(y))$ at the localization on the brane is decided by $F(\phi(y))$. This point will be more clear when it is considered a specific Yukawa coupling. Therefore, the behavior of $F(\phi(y))$ plays a leading role for the fermion localization on the brane \cite{yo}. Having set up the two essential conditions for the problem of fermion localization on the brane, we are now in a position to choice some specific forms for Yukawa couplings.

\subsection{Zero mode and fermion localization}

From now on, we mainly consider the simplest case $F(\phi)=\phi$. First, we consider the
normalizable problem of the solution. In this case, we only need to consider the asymptotic behavior of the integrand in (\ref{cono}). It becomes
\begin{equation}\label{in}
    I\rightarrow\mathrm{exp}\left[ -A(\pm\infty)-2\eta\int^{y}_{0}\phi(y^{\prime})dy^{\prime} \right]\,.
\end{equation}

\noindent By the analysis from eq. (\ref{em2b}), we obtain that $A(\pm\infty)\rightarrow-\sqrt{|V(\pm v)|/3}\,|\,y|$. For the integral $\int dy\phi$, we only need to consider the asymptotic behavior of $\phi$ for $y\rightarrow\pm\infty$ \cite{liu} and as $\phi(\pm\infty)=\pm v$ the equation (\ref{in}) becomes
\begin{equation}\label{int}
    I\rightarrow\mathrm{exp}\left[ -2\left( \eta\,v-\sqrt{|V(\pm v)|/12} \right)|\,y| \right]\,.
\end{equation}

\begin{figure}[ht] 
       \begin{minipage}[b]{0.40 \linewidth}
           \fbox{\includegraphics[width=\linewidth]{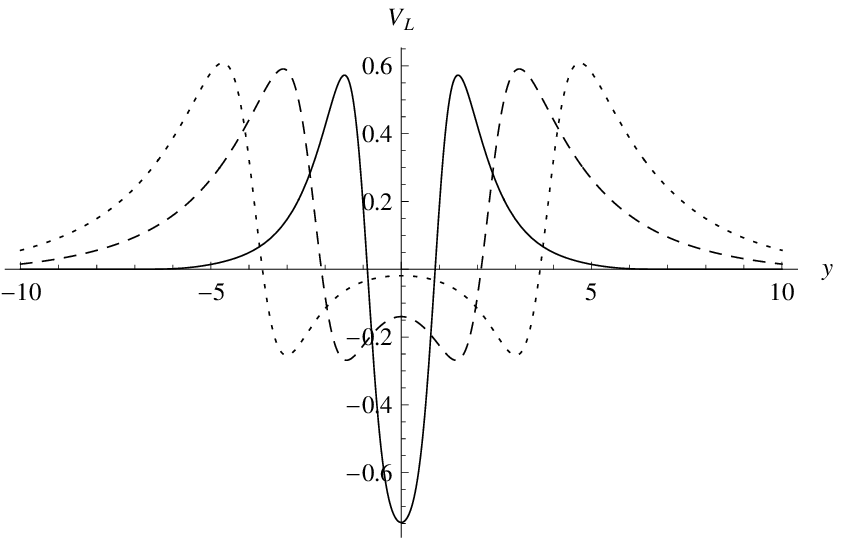}}\\
          \end{minipage}\hfill
       \begin{minipage}[b]{0.40 \linewidth}
           \fbox{\includegraphics[width=\linewidth]{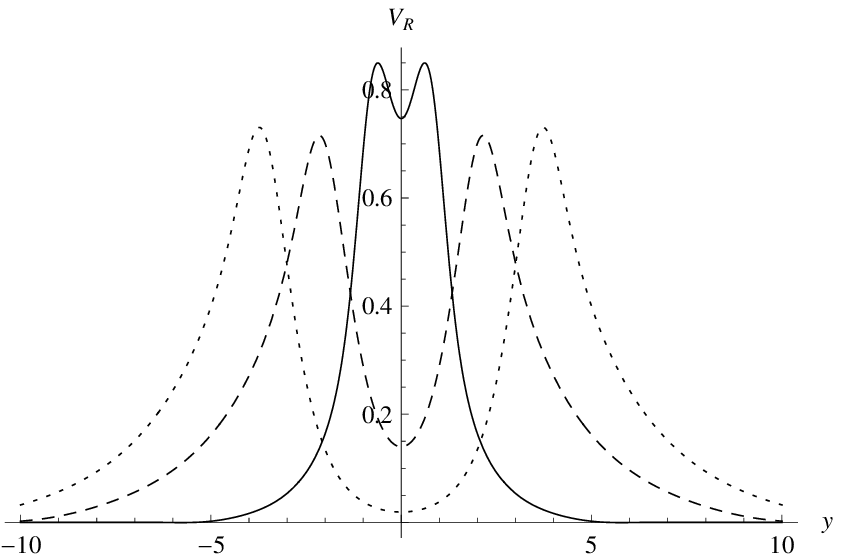}}\\
           \end{minipage}
       \caption{Potential profile: (a) $(V_{L}(y))_{A}$ (left) and (b) $(V_{R}(y))_{A}$ (right) for $\eta=1$, $b=2$,  $c=1$, $a=0$ (thin line), $a=0.8$ (dashed line) and $a=0.836$ (dotted line) }\label{pe}
   \end{figure}

\begin{figure}[ht] 
       \begin{minipage}[b]{0.40 \linewidth}
           \fbox{\includegraphics[width=\linewidth]{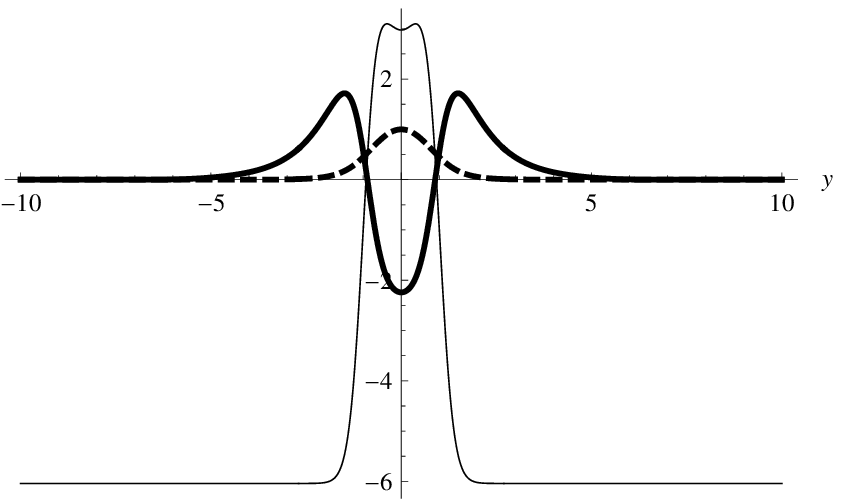}}\\
          \end{minipage}\hfill
       \begin{minipage}[b]{0.40 \linewidth}
           \fbox{\includegraphics[width=\linewidth]{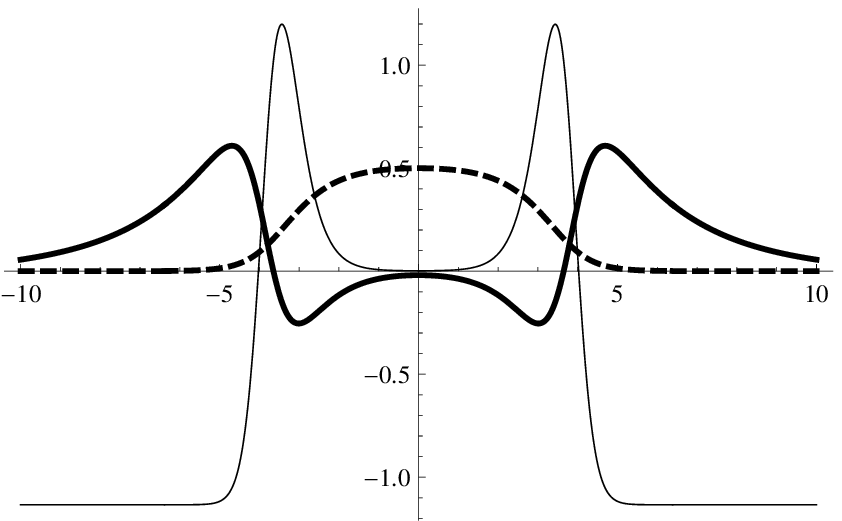}}\\
           \end{minipage}
       \caption{The profiles of the Ricci scalar (thin line), $V_{L}(y)$ (thick line) and $|\,L_{0}|^{2}$ (dashed line) for $\eta=1$, $b=2$, $c=1$; (a) $a=0$ (left) and (b) $a=0.836$ (right).}\label{a}
   \end{figure}

\noindent This result clearly shows that the zero mode of the left-handed fermions is normalized only for $\eta>\frac{1}{v}\,\sqrt{\frac{|V(\pm v)|}{12}}$. Now, under the change $\eta\rightarrow-\eta$ ($L_{0}\rightarrow R_{0}$) we obtain that the right-handed fermions can not be a normalizable zero mode. The shape of the potentials $V_{L}$ and $V_{R}$ are shown in Fig. (\ref{pe}) for some values of $a$. Figure \ref{pe}(a) shows that the effective potential $V_{L}$, is indeed a volcano-like potential for $a=0$. As $a$ increases the well structure of $V_{L}$ gets a double well. Figure \ref{pe}(b) shows that the potential $V_{R}$ has also a well structure, but the minimum of $V_{R}$ is always positive, therefore the potential does not support a zero mode. The shapes of the matter energy density, $V_{L}$ potential and $|\,L_{0}|^{2}$ are shown in Fig. \ref{a}. The Fig. \ref{a}(a) ($a=0$) shows that the zero mode is localized on the brane. On the other hand, Fig. \ref{a}(b) ($a=0.836$) clearly shows that the normalizable zero mode is localized between the two sub-branes, as a consequence the zero mode is not localized on the brane. Therefore, we can conclude that the zero mode of the left-handed fermions is localized on the brane only as $0\leq a \approx a_{*}$.

\section{Conclusions}

We have reinvestigated the effects of the variation of the mass parameter $a$ on the thick branes as well as the localization of fermions. We showed that the variation of $a$ is associated to the phenomenon of brane splitting, therefore the brane does not get thicker with the increases of the value of $a$, as argued in Ref. \cite{zhao}. We can conclude that the appearance of two sub-branes is associated to phase transition for $a=a_{*}$ (a disordered phase between two ordered phases). Also, we showed that the value of the matter energy depends on the asymptotic behavior of the warp factor. From the static equation of motion we have analyzed the asymptotic behavior of $A(y)$ and showed that the zero mode of the left-handed fermions for the simplest Yukawa coupling $\eta\bar{\Psi}\phi\Psi$ is normalizable under the condition $\eta>\frac{1}{v}\,\sqrt{\frac{|V(\pm v)|}{12}}$ and it can be trapped on the brane only for $0\leq a \approx a_{*}$, because the zero mode has a single-peak at the localization of the brane. We also showed that as $a\rightarrow a_{*}$ the zero mode has a single-peak between the two sub-branes and as a consequence the normalizable zero mode is not localized on the brane. Therefore, the brane with a big value of $a$ would not trap fermions more efficiently, in opposition to what was adverted in Ref. \cite{zhao}. This work completes and revises the analyzing of the research in Ref. \cite{zhao}, because in that work does not analyze the zero mode in full detail and contain some misconceptions.

Additionally, we showed that the change of variable $dz=\mathrm{e}^{-A(y)}dy$ leads to a non conserved current, because the curved-space gamma matrices are not covariantly constant. An interesting issue will be investigate the effects of non-conserved current on resonances modes and bear out the main conclusion of Ref. \cite{zhao}.

\begin{acknowledgments}
This work was supported by means of funds provided by CAPES.
\end{acknowledgments}

\end{document}